\documentclass[
aps,%
12pt,%
final,%
notitlepage,%
oneside,%
onecolumn,%
nobibnotes,%
nofootinbib,% In the current version of REVTeX, when this option is on,
superscriptaddress,%
noshowpacs,%
centertags]%
{revtex4}
\usepackage[cp1251]{inputenc}
\usepackage[T2A]{fontenc}
\usepackage[english,russian]{babel}
\usepackage{amsfonts}\usepackage{amsbsy}
\usepackage{color}
\usepackage{amsfonts}
\usepackage{amsbsy}
\usepackage{mathrsfs}
\usepackage{graphicx}
\newcommand{\nl}{\nonumber\\ }
\newcommand{\pd}{\partial}

\def\bc{\begin{center}}
\def\ec{\end{center}}
\def\om{\omega}
\def\be{\begin{eqnarray}}
\def\ee{\end{eqnarray}}
\def\prt{\partial}
\def\lsim{\stackrel{\scriptstyle <}{\phantom{}_{\sim}}}
\def\gsim{\stackrel{\scriptstyle >}{\phantom{}_{\sim}}}

\begin{document}
\title{%YaF%\vspace*{-15mm}
Viscosity of   hadron matter  within relativistic mean-field based model with
scaled hadron masses and couplings }
\author{\firstname{A.S.~Khvorostukhin}}
%\thanks{e-mail: Y.Ivanov@gsi.de}
\email[]{hvorost@theor.jinr.ru}
%\homepage[]{Your web page}
%\thanks{}
\altaffiliation{Joint Institute for Nuclear Research,
%141980
Dubna, Russia}
%\noaffiliation
\affiliation{Institute of Applied Physics, Moldova Academy of Science,
% MD-2028
Kishineu, Moldova}
\author{\firstname{V.D.~Toneev}}
%\thanks{e-mail: russ@ru.net}
\email[]{toneev@theor.jinr.ru}
%\homepage[]{Your web page}
%\thanks{}
\altaffiliation{Joint Institute for Nuclear Research,
% 141980 123182
Dubna, Russia}
%\noaffiliation
\affiliation{Gesellschaft f\"ur Schwerionenforschung mbH,
%Planckstr.$\!$ 1, 64291
Darmstadt, Germany}
\author{\firstname{D.N.~Voskresensky}}
%\thanks{e-mail: russ@ru.net}
\email[]{voskre@pisem.net}
%\homepage[]{Your web page}
%\thanks{}
\altaffiliation{Moscow Engineering Physical Institute,
% 123182
Moscow, Russia}
%\noaffiliation
\affiliation{Gesellschaft f\"ur Schwerionenforschung mbH,
%Planckstr.$\!$ 1, 64291
Darmstadt, Germany}
\begin{abstract}

The shear ($\eta$) and bulk ($\zeta$) viscosities are calculated
in a quasiparticle relaxation time approximation for a  hadron
matter described within the relativistic mean-field based model
with scaled hadron masses and couplings.  Comparison with results
of other models is presented. We demonstrate that a small value of
the shear viscosity to entropy density ratio required for
explaining a large elliptic flow observed at RHIC may be
reached in the hadron phase. Relatively large values of the bulk 
viscosity are noted in the case of a baryon enriched matter.
%\keywords{relativistic mean-field, viscosity coefficients, relativistic heavy-ion collisions}
\end{abstract}
\maketitle
%\today

\section{Introduction}

 In
the past, transport coefficients for the nuclear matter were
studied in \cite{AW73,Gal79,Dan84,HM93}.  Recently, the interest
in the transport coefficient issue  has sharply been increased in
heavy-ion collision physics, see review-article~\cite{K08}. Values
of the elliptic flow $v_2$
  observed at  RHIC~\cite{RHIC-v2} proved to be larger than at  SPS. This
finding is interpreted as that a quark-gluon plasma (QGP) created at
RHIC behaves as a nearly perfect fluid with a small value of the
shear viscosity-to-entropy density ratio, $\eta/s$. The latter
statement was confirmed by non-ideal hydrodynamic analysis of
these data~\cite{Tea03}. Thereby, it was
claimed~\cite{PSS01,PC05,Sh05} that a new state produced at high
temperatures is most likely not a weakly interacting QGP, as it
was originally assumed, but a strongly interacting QGP. The
interest was also supported by a new theoretical perspective,
namely, ${\cal N}=4$ supersymmetric Yang-Mills gauge theory using
the Anti de-Sitter space/Conformal Field Theory (AdS/CFT) duality
conjecture. Calculations in this strongly coupled theory
demonstrate that there is  minimum in the $\eta/s$
ratio~\cite{KSS03}:
%\begin{equation}
${\eta}/{s}\approx{1}/{(4\pi)}.$
% \label{KSS}
% \end{equation}
It was thereby conjectured that this relation
%(\ref{KSS})
is in fact a lower bound for the specific shear viscosity in all
systems~\cite{K08} and that the minimum is reached in the
hadron-quark  transition critical point (at $T=T_c$).

In this paper,  we continue investigation of the shear and bulk
viscosities performed in Ref. \cite{SR08}  within the
quasiparticle model in the relaxation time approximation. We
describe the hadron phase ($T<T_c$) in terms of the quasiparticle
relativistic mean-field-based model with the scaling hadron masses
and couplings (SHMC) been successfully applied to the description
of heavy ion collision reactions ~\cite{KTV07,KTV08}, see sect.2.
Then in sect.3 we calculate the shear and bulk viscosities and
compare our results with results of previous works. In sect. 4 we
formulate our conclusions.

\section{Description of hadron matter in the SHMC model}\label{Rem}
\subsection{Formulation of the model}

 Within our relativistic mean-field SHMC model~\cite{KTV07,KTV08}
we present the Lagrangian density of the hadronic matter as a sum
of several terms:
 \be\label{math}  \mathcal{L}=\mathcal{L}_{\rm bar}+\mathcal{L}_
 {\rm MF}+\mathcal{L}_{\rm ex}~.
 \ee
The  Lagrangian density of the baryon component interacting via
$\sigma,\omega$ mean fields  is as follows:
 \be
 \mathcal{L}_{\rm bar} &=& \sum_{b\in {\rm \{ bar \} }} \left[ i\bar \Psi_b\,
\Big(\prt_\mu +i\,g_{\om b} \,{\chi}_\om
 \ \om_\mu \Big) \gamma^\mu\, \Psi_b -m_b^*\,
\bar\Psi_b\,\Psi_b
 \right] .
 \label{lagNn}
 \ee
  The considered baryon set is $\{b\}=N(938)$, $\Delta (1232)$,
$\Lambda (1116)$, $\Sigma (1193) $, $\Xi (1318)$, $\Sigma^*
(1385)$, $\Xi^* (1530)$, and $\Omega (1672)$, including
antiparticles. The used $\sigma$-field dependent effective masses
of baryons are~\cite{KTV07,KTV08,KV04}
\be \label{bar-m}
{m_b^*}/{m_b}=\Phi_b(\chi_\sigma  \sigma)= 1 -g_{\sigma b} \
\chi_{\sigma} \ \sigma /m_b \,, \; b\in\{b\}~.
 \ee
 In Eqs.
(\ref{lagNn}), (\ref{bar-m})  $g_{\sigma b}$ and $g_{\om b}$  are
coupling constants and $\chi_\sigma (\sigma)$, $\chi_\om (\sigma)$
are coupling  scaling functions.

The $\sigma$-, $\omega$-meson mean field contribution is given by
\begin{eqnarray} \mathcal{L}_{\rm
MF}&=&\frac{\prt^\mu \sigma \ \prt_\mu
\sigma}{2}-\frac{m_\sigma^{*2}\,
\sigma^2}{2}-{U}(\chi_{\sigma}\sigma)
-\frac{\omega_{\mu\nu}\,\omega^{\mu\nu}}{4} +\frac{m_\om^{*2}\,
\om_\mu\om^\mu}{2}~,\\
 \omega_{\mu\nu}\,&=&\partial_\mu \om_\nu -\partial_\nu \om_\mu ~,\quad U(\chi_{\sigma}\sigma)
 =m_N^4 (\frac{b}{3}\,f^3 +\frac{c}{4}\,f^4 ),
 \quad  f=g_{\sigma N} \ \chi_\sigma \ \sigma/m_N\,.\nonumber
\end{eqnarray}
 There exist only $\sigma$ and $\om_0$ mean field solutions of
equations of motion. The mass terms of the mean fields are \be
\label{bar-m1} {m_m^*}/{m_m}&=&|\Phi_m (\chi_\sigma \sigma)|\,,
\quad \{m\}=\sigma,\om\,. \ee

The dimensionless scaling functions $\Phi_b$ and $\Phi_m$, as well
as the coupling scaling functions $\chi_m$, depend on the scalar
field in combination $\chi_\sigma(\sigma) \ \sigma$.
 Following
\cite{KV04} we assume approximate validity of the Brown-Rho
scaling ansatz in the simplest form
\be \label{Br-sc}\Phi =\Phi_N
=\Phi_\sigma =\Phi_\om =\Phi_\rho =
 1-f .
 \ee
The third term in  the Lagrangian density (\ref{math}) includes
meson quasiparticle excitations:
$\pi;K,\bar{K}; \eta (547);
\sigma',\omega',\rho';K^{*\pm,0}(892),\eta'(958),\phi(1020).$
 The choice of parameters and other details of the SHMC model
can be found in~\cite{KTV07,KTV08}.

\subsection{Thermodynamical quantities}
\begin{figure}[thb]
\includegraphics[width=130mm,clip]{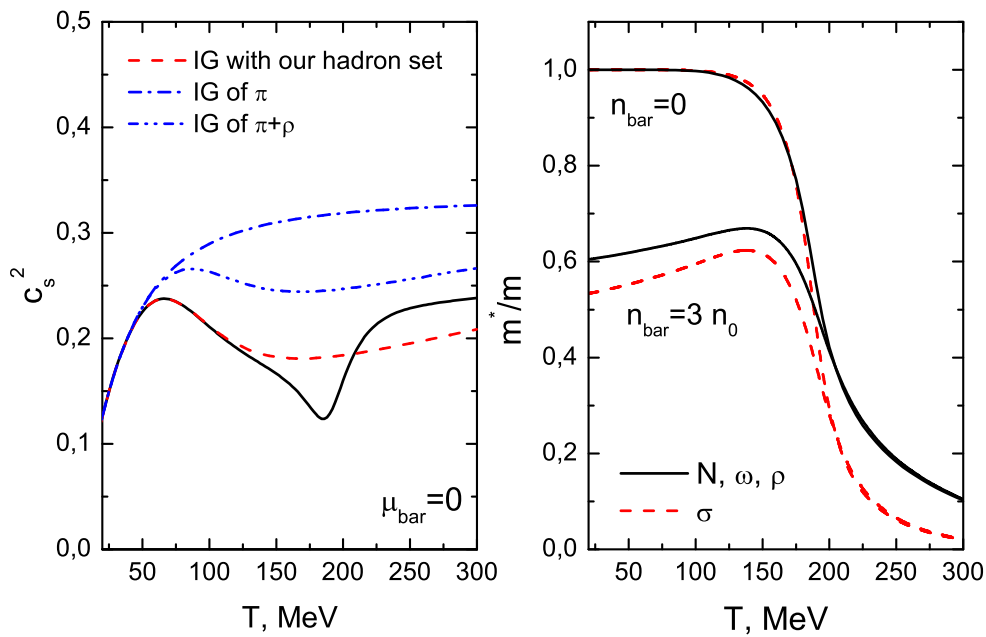}
\caption{ Left panel: The sound velocity squared  in hadron matter
as function of the temperature at zero baryon chemical potential.
Solid line -- calculation within the SHMC model. Other notations
are given in the legend.
Right panel: The temperature dependence of effective  masses of
the nucleon, $\omega$ and $\rho$ excitations (solid line) and of
the $\sigma$-meson excitation (dashed line) calculated within the
SHMC model for two values of  the baryon density.
 }
 \label{s-vel_h}
\end{figure}

Within SHMC model we calculate different thermodynamical
quantities in thermal equilibrium hadron matter at fixed
temperature $T$ and baryon chemical potential $\mu_{\rm bar}$. In
Fig.~\ref{s-vel_h} (left panel) we show the square of the sound
velocity $c_s^2=dP/d\varepsilon$ ($P$ is pressure, $\varepsilon$
is energy density), as function of temperature at zero baryon
chemical potential, $\mu_{\rm bar} =0$, for the SHMC model (solid
line) and compare this result with that for the ideal gas (IG)
model with the same hadron
set as in the SHMC model (long-dashed line), for the $\pi +\rho$
mixture (dash-double dot) and for purely pion system
(dash-dotted).
 As is seen from this figure, for the purely pion IG the  $c_s^2$
monotonously increases with  increase of the temperature
approaching the ultrarelativistic  limit $c_s^2=1/3$ at high
temperatures.  For the pion-rho meson mixture, the $c_s^2$
exhibits a shallow minimum at $T\sim 170$ MeV. The minimum (in the
same temperature region) becomes more pronounced for
multi-component systems (see dash curve). At $T\lsim$ 50 MeV the
pion contribution is a dominant one, thereby all curves
coincide\footnote{Note that within the SHMC model pions are
treated as an ideal gas of free particles}. The curves for the
SHMC model and the IG model calculated with the same hadron set
coincide for $T\lsim$ 100 MeV. At $T>$ 50 MeV  heavier mesons
start to contribute that slows down the growth of pressure and
then results in significant decrease of $c_s^2$, contrary  to the
case of the one-component pion gas.
 Within the SHMC model
$c_s^2$ gets pronounced minimum at $T\simeq$ 180 MeV caused by a
sharp decrease of the in-medium hadron masses at these
temperatures (see the right panel in Fig.~\ref{s-vel_h}, where
effective masses of the nucleon, $\omega$, $\rho$ and $\sigma$
excitations are presented). The minimum of the sound velocity (at
$T= T_c \simeq$ 180 MeV) can be associated with a kind of phase
transition, e.g. with the hadron-QGP cross-over, as it  follows
from the detailed analysis of the lattice data, see
\cite{karsch06}.

 Note that in the
Hagedorn-gas model~\cite{CCMS09}  (for the Hagedorn mass $m\to
\infty$) one gets $c_s^2\to 0$ at $T=T_c$, whereas in the
mass-truncated Hagedorn-gas model  the behavior very close to that
we have in case of the IG model is  observed.

\begin{figure}[thb]
 \hspace*{10mm}
\includegraphics[width=65mm,clip]{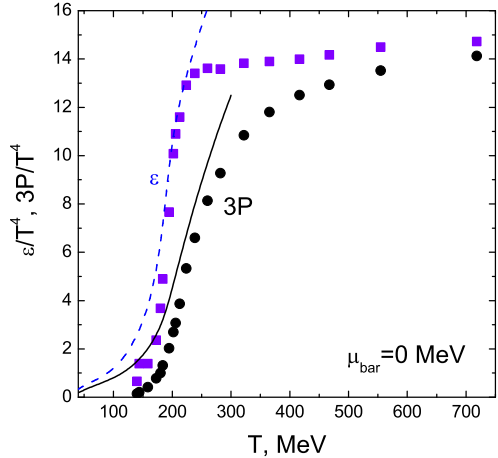}
\caption{ The reduced triple pressure and the energy density  at
$\mu_{\rm bar}=0$.  Points are QCD lattice result \cite{karsch06}.
The hadronic SHMC results   are plotted by solid and dash lines,
respectively.}
 \label{epT4}
\end{figure}

In Fig.~\ref{epT4} we present the lattice data for the reduced
energy density and the pressure together with our SHMC model
results. Following ~\cite{KTV08} we use   suppressed coupling
constants $g_{\sigma b}$ (except for nucleons). This  guarantees
that even above $T_c$ up to the temperature $T\sim $ 220 MeV
 the EoS computed in the SHMC model 
is in agreement with the lattice data for the pressure and energy 
density. At higher temperatures
the SHMC model requires additional modifications,  although
in reality the quark-gluon degrees of freedom should be taken into
account already for $T>T_c \sim 180$ MeV.

\section{Shear and bulk viscosities of the SHMC
model}\label{ViscH}

\subsection{Collisional viscosity, derivation of equations}
 Sasaki and Redlich \cite{SR08} derived expressions for the
shear and bulk viscosities  in the case when the quasiparticle
spectrum is given by
%\be
$E(\vec{p})=\sqrt{\vec{p}^{\,\,2} +m^{*\,2}(T,\mu)}~.$
% \label{qe}
%\ee
We perform a similar derivation, but in the presence of mean
fields. In the latter case one should additionally take into
account that quasiparticle distributions depend on
the mean fields.

 In order to calculate viscosity coefficients one needs an expression
for spatial components of the energy momentum tensor
corresponding to the Lagrangian density (\ref{math}):
\begin{eqnarray}
T^{ij}=T^{ij}_{\rm MF}+\sum_{b\in \{\rm bar\}}T^{ij}_b +
\sum_{{\rm bos}\in \{\rm ex\}}T^{ij}_{\rm bos}, \label{tmunuT}
\end{eqnarray}
where $i,j =1,2,3$ and the mean-field contribution is as follows
\begin{eqnarray}
&&T^{ij}_{\rm MF}=\partial^{i}\sigma \ \partial^{j}\sigma
-\partial^{i}\om_0 \ \partial^{j}\om_0 \nonumber\\ &&+
\left(\frac12\left[\partial^{i}\sigma \ \partial^{j}\sigma
-\partial^{i}\om_0 \ \partial^{j}\om_0 +{m^*_\sigma}^2 \ \sigma^2
-{m^*_\omega}^2 \ \omega_0^2\right]+U(\sigma) \right)g^{ij}
\end{eqnarray}
with $m^*_\sigma$ and $m^*_\omega$  given by Eq.~(\ref{bar-m1}).

The quasiparticle (fermion and boson excitation) contribution is
given by
\begin{eqnarray}\label{Tmunua}
T^{ij}_a =\int d\Gamma \frac{p_a^i  p_a^j}{E_a}F_a ~, \; \; \;
a\in ({\rm bos., bar}),\quad E_a =\sqrt{\vec{p}^{\,\,2}
+m^{*\,2}}~,\quad\Gamma =
 \nu_a \frac{d^3p_a}{(2\pi)^3},
\end{eqnarray}
where $\nu_a$ is the degeneracy factor.

The  quasiparticle distribution function $F_b$ for baryon
components in the presence of mean fields fulfills the Boltzmann
kinetic equation \cite{SCFNW},
\begin{eqnarray}\label{Boltz1}
\left(p^{\mu}_b\partial_{\mu}-g_{\om
b}p_{\mu}\om^{\mu\nu}\frac{\pd }{\pd
{p}^{\nu}_{b}}+m_b^{*}\partial^{\nu}m_b^{*}\frac{\pd }{\pd
{p}^{\nu}_{b}}\right)\widetilde{F}_b=St \widetilde{F}_b ;
\end{eqnarray}
with $\widetilde{F}_b  ({p}_b ,x_b) =\delta (p_b^2
-m_b^{*\,2})F_b (\vec{p}_b ,x_b) $.

   The local equilibrium boson or baryon distribution is given as follows:
 \be
 F^{\rm loc.eq.}_a (\vec{p}_a ,x_a )=\left[e^{(E_a -\vec{p}_a \vec{u}-
\mu_a + t_a^{\rm vec}X_a^0 )/T}\pm 1 \right]^{-1},\quad X_a^0 =
g_{\om a} \ \chi_{\om} \ \om_0 ,
 \label{leqdf}
 \ee
where we suppressed $\vec{u}^{\,2}$ terms for $|\vec{u}|\ll 1$.
Here the upper sign ($+$) is for fermions and ($-$) is
 for bosons, and the vector particle charge is
 $ t_a^{\rm vec}=\pm 1$ or $0$; $g_{\om a}\neq 0$ only for $a\in \rm bar$ 
 in our model.
Considering only slightly inhomogeneous  solutions and using
$|\vec{u}|\ll 1$ we may drop the terms $\propto \vec{u}^2$ and
$\propto \vec{u}\nabla \om_0$ in the kinetic equation
(\ref{Boltz1}). Then kinetic equations for boson and baryon
components acquire ordinary quasiparticle form
\begin{eqnarray}\label{Boltz}
\frac{\pd F_a}{\pd t}+\frac{\pd E_a}{\pd \vec{p}_a}\frac{\pd
F_a}{\pd\vec{r}_a}-\frac{\pd E_a}{\pd \vec{r}_a}\frac{\pd
F_a}{\pd\vec{p}_a}=\frac{p^{\mu}_a}{E_a} \ \frac{\partial
F_a}{\partial x^{\mu}_a}=St F_a ,
\end{eqnarray}
 where  $p_a^\mu = (E_a (\vec{p}_a ,\vec{r}_a ,\sigma,\om ),\vec{ p}_a)$.
 %, $u_{\mu}$ is the flow 4-velocity.
  We used  that
$\pd E_a /\pd \vec{p}_a=\vec{p}_a/ E_a$.
 Since  calculating the viscosity,  we need only terms with
velocity gradients, we further put $\pd E_a/\pd \vec{r}_a=(\pd
E_a/\pd \mu_a)\vec{\nabla}_a\mu_a +(\pd E_a/\pd T)\vec{\nabla}_a T
=0$.

 In the relaxation time approximation
 \be
 \label{st}
 St F_a =-\delta F_a /\tau_a , \quad\delta F_a = F_a -F_a^{\rm loc.eq.} .
 \ee
 Here $\tau_a$ denotes the relaxation time of the given
species. Generally, it depends on the quasiparticle momentum
$\vec{p}_a$ and the quasiparticle energy $E_a (\vec{p}_a)$.

 The averaged  partial relaxation time ${\tilde\tau}_a$ is related to
 the cross section   as
 \be
  {\tilde \tau}^{-1}_a (T,\mu )
=\sum_{a^{'}} n_{a^{'}} (T,\mu)\left<v_{aa^{'}}
\sigma^t_{aa^{'}}(v_{aa^{'}}) \right>,
 \label{tau}
 \ee
where $n_{a^{'}}$ is the density of $a^{'}$-species,
$\sigma^t_{aa^{'}}=\int d\cos \theta \ d\sigma(aa^{'}\to
aa^{'})/d\cos \theta \ (1-\cos \theta) $ is the transport cross
section, in general,  accounting for in-medium effects and
$v_{aa^{'}}$ is the relative velocity of two colliding particles
$a$ and $a^{'}$ in  case of binary collisions. Angular brackets
denote a quantum mechanical statistical average over an
equilibrated system. In reality, the cross sections entering the
collision integral and the corresponding relaxation time $\tau_a$
in (\ref{st}) may essentially depend on the particle momentum.
Thus, averaged values ${\tilde\tau}_a^{-1}$ given by Eq.
(\ref{tau}) yield only a rough estimate for the values
${\tau}^{-1}_a$ which we actually need for calculation of
viscosity coefficients, see below Eqs. (\ref{shear}) and
(\ref{bulk}).

 In the relaxation time approximation from
 Eqs. (\ref{Boltz}), (\ref{st}) we obtain
\be\label{deltaF}
 \delta F_a =-\frac{\tau_a}{E_a } \ p^{\mu}_a  \ \frac{\partial
 F_a^{\rm loc.eq.}}{\partial
x^{\mu}_a},
 \ee
 and then the variation of the energy-momentum tensor
(\ref{tmunuT}) becomes:
 \be
 \label{varenmom} \delta T^{ij} =-\sum_a\int
d\Gamma \left\{\tau_a\frac{p_a^i  p_a^j}{E_a^2} \ p_a^\mu\pd_\mu
F_a \right\}_{\rm loc.eq.} +\delta\sigma \
\left\{\frac{\partial T^{ij}}{\partial\sigma}\right\}_{\rm
loc.eq.} +\delta\om_0 \ \left\{\frac{\partial
T^{ij}}{\partial\om_0}\right\}_{\rm loc.eq.}~.
 \ee
Considering small deviations from the local equilibrium, we may
keep in (\ref{varenmom}) only first-order derivative quasiparticle
terms $\propto \partial_{i}$, thus neglecting mean-field
contributions $\propto \partial_i \sigma \ \partial^j\sigma$ and
$\propto
\partial_i \om_0 \ \partial^j\om_0$.

The shear and bulk viscosities are as follows expressed through
the variation of the energy-momentum tensor:
 \be
 \label{vis}
\delta T_{ij}=-\zeta \ \delta_{ij}{\vec{\nabla}}\cdot\vec{u}-\eta
\ W_{ij},\quad
W_{kl}= \pd_ku_l+\pd_lu_k-\frac23\delta_{kl} \ \pd_iu^i~.
 \ee
 To find the shear viscosity, we put $i\neq j$ in (\ref{vis})
 and use that in this case the variation of the second and
third terms in (\ref{varenmom}) yields zero after integration over
angles. To find the bulk viscosity, we  substitute $i=j$ in
(\ref{vis}) and use that $T^{ii}_{\rm eq}=3P_{\rm eq}$.  We put
$\vec{u}=0$ in final expressions but retain gradients of the
velocity.

Taking derivatives  $\partial F_a^{\rm loc.eq.}/\partial
x^{\mu}_a$ in Eq. (\ref{deltaF}) we find the variation of the
total energy-momentum tensor as the function of derivatives of the
velocity
 \be
 \label{deltaTij}
\delta T^{ij}&=&\sum_{a}\int d\Gamma \ \frac{p^i_a
p^j_a}{TE_a}\,\tau_a \ F_a^{\rm eq}(1\mp  F_a^{\rm eq})\,q_a (\vec
p\,;T,\mu_{\rm bar} ,\mu_{\rm str})~
 \ee
 with
 \be
 q_a (\vec p\,;T,\mu_{\rm
bar} ,\mu_{\rm str} )=\pd_ku_l \ \delta_{kl} \ Q_a
-\frac{p_kp_l}{2E_a} \ W_{kl},
 \ee
\begin{eqnarray}\label{Qa} &Q_a =
-\left\{\frac{\vec{p}_a^{\,2}}{3E_a}+\left(\frac{\pd P}{\pd n_{\rm
bar}}\right)_{\epsilon,n_{\rm str}}\left[\frac{\pd (E_a
+X^0_a)}{\pd\mu_{\rm bar}}-t_b^{\rm bar}
\right]\right.\nl&\left.+\left(\frac{\pd P}{\pd n_{\rm str}
}\right)_{\epsilon,n_{\rm bar}}\left[\frac{\pd (E_a
+X^0_a)}{\pd\mu_{\rm str}}-t_a^{\rm str}\right]-\left(\frac{\pd
P}{\pd \epsilon}\right)_{n_{\rm bar} ,n_{\rm str} }\times\right.\\
&\left.\times\left[E_a +X^0_a -T\frac{\pd (E_a +X^0_a )}{\pd
T}-\mu_{\rm bar}\frac{\pd (E_a +X^0_a)}{\pd \mu_{\rm bar}
}-\mu_{\rm str} \frac{\pd (E_a +X^0_a)}{\pd \mu_{\rm str}
}\right]\right\}.\nonumber
\end{eqnarray}
 Finally, we obtain
expressions for the shear viscosity
\begin{eqnarray}\label{shear}
\eta&=\frac{1 }{15T}\sum_{a}\int d\Gamma\,\tau_a
\frac{\vec{p}^{\,4}_a}{E^2_a}\,\left[ F^{\rm eq}_a \; (1\mp F^{\rm
eq}_a )\right],
\end{eqnarray}
and for the bulk viscosity
\begin{eqnarray}
\label{bulk}
\zeta&=-\frac{1}{3T}\sum_{a} \int\mathrm d\Gamma\,\tau_a \;
\frac{\vec{p}^{\,2}_a}{E_a} \; F^{\rm eq}_a \;
\left(1\mp F^{\rm eq}_a\right)Q_a .
\end{eqnarray}
 At vanishing mean fields our results are reduced to those derived in 
 Ref. \cite{SR08}.

\subsection{Collisional viscosity  in baryon-less matter}

 In the relaxation time
approximation both shear and bulk viscosities for a component
"$a$" \ depend on its relaxation (collisional) time $\tau_a$ which
should be parameterized or calculated independently. Therefore to
diminish this uncertainty it is legitimate at first to find the
reduced kinetic coefficients (per unit relaxation time, assuming
$\tau =const$, i.e. $\tau ={\tilde\tau}$).

\begin{figure}[h]
 \hspace*{2mm}
\includegraphics[width=120mm,clip]{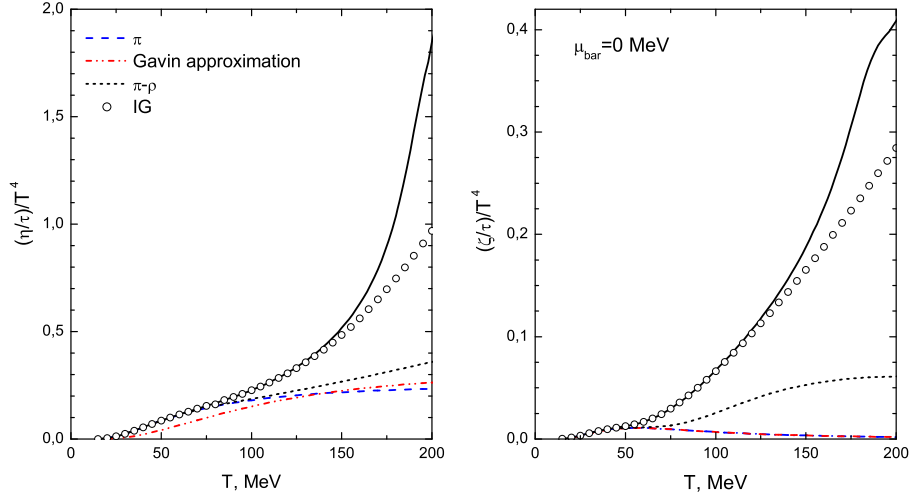}
\caption{  The reduced (per unit relaxation time) $T^4$ scaled
shear (left panel) and bulk (right panel) viscosities as function
of the temperature calculated within the SHMC model  (solid lines)
for the baryon-less matter, $\mu_{\rm bar}=$0. Results are
compared with those for the massive pion gas (dashed lines),
$\pi-\rho$ mixture (short dashed
 lines)  and with those for the  massless pion gas
(the Gavin approximation~\cite{G85}, dot-dashed line), as well as
for the IG model (open dots) with the same  set of species as in
the SHMC model.
 }
 \label{Gav}
\end{figure}

In Fig.~\ref{Gav} we demonstrate  results of various calculations
for the reduced shear (left panel) and  bulk  (right panel)
viscosities  scaled by the $1/T^4$ factor  at $\mu_{\rm bar}=0$.
As we see from the figure, the reduced shear viscosity of the
massive pion gas (dashed line) becomes approximately  proportional
to $T^4$ for $T\gsim$ 100 MeV. Naturally,  this result is close to
that obtained in the Gavin approximation~\cite{G85}
(dashed-double-dotted line in Fig.~\ref{Gav}).   The $T^4$ scaling
is violated for the $\pi-\rho$ gas in the temperature interval
under consideration because the $\rho$ mass is not negligible even
at $T\sim $ 200 MeV. For $\zeta$ the approximate $1/T^4$ scaling
property holds for the massive pion-rho gas at $T\gsim$ 150
MeV. Note that $\zeta=$0 for the  gas of free massless pions since
$c_s^2=1/3$ in this case. For the massive pion gas $\zeta/T^4$
decreases already  for
 $T>60$ MeV reaching zero at large $T$ similar to the massless gas.
The reduced shear and bulk  viscosities of a multicomponent system
calculated in  our SHMC model (solid lines) and in the IG model
with the same hadron set (open dots) do not fulfill the $T^4$
scaling law. These models include  large set of hadrons, due to
that with the temperature increase  the
reduced shear and bulk viscosities    become significantly higher
 than those for the pion gas and the
pion-rho gas models. An additional increase of the reduced
viscosity within SHMC model originates from  significant mass
decrease at temperatures near the critical temperature. The bulk
viscosity of a single-component pion system drops to zero both at
low and high temperatures and in the whole temperature interval
$\zeta<<\eta$, that is frequently used as an argument for
neglecting the bulk viscosity effects. However, the statement does
not hold anymore for  mixture of many species. For example, at
$T\sim$150 MeV the $\eta/\zeta$ ratio is only about 3 in case
of the IG and SHMC models.  Thus the
bulk viscosity effects can play a role in the description of  the
hadronic stage at high collision energies, like at RHIC.
Moreover, the bulk viscosity can be responsible  for such
important effect as flow anisotropy.

\subsection{Collisional viscosity  in baryon enriched matter}

\begin{figure}[thb]
 \hspace*{2mm}
\includegraphics[width=65mm,clip]{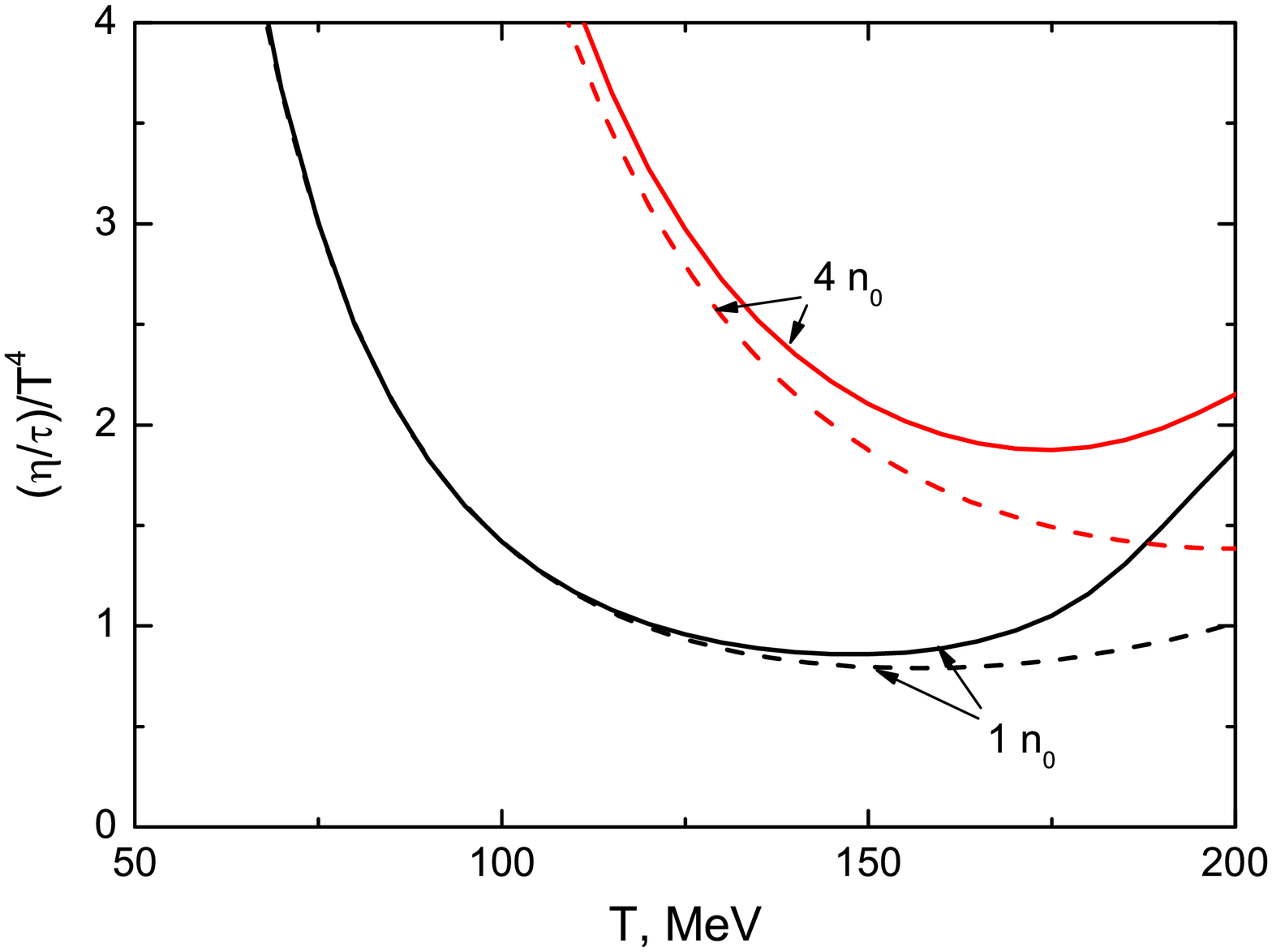}
\includegraphics[width=65mm,clip]{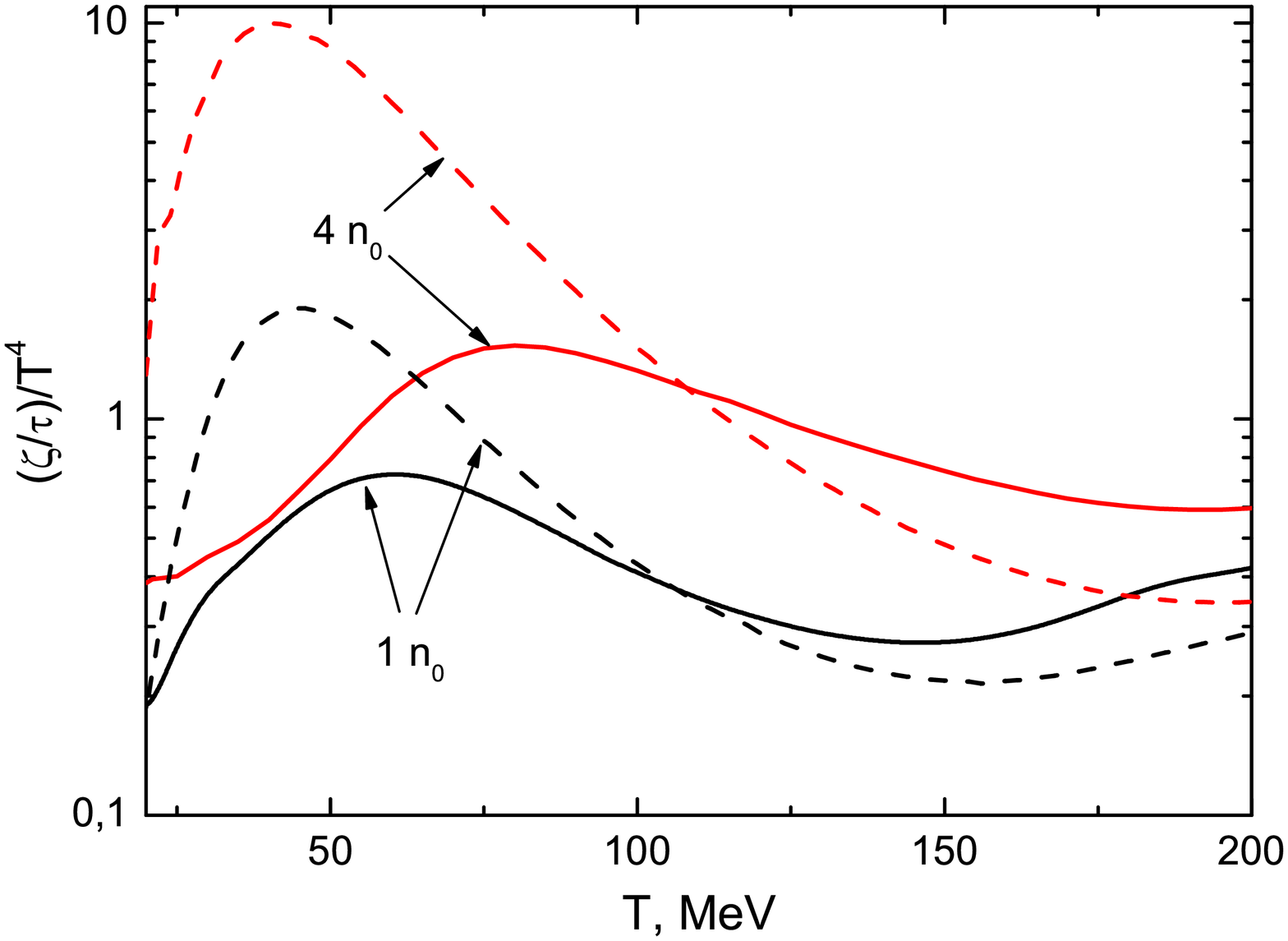}
\caption{The SHMC  model predictions of the $T^4$ scaled
temperature dependence of the reduced shear (left panel) and bulk
(right panel) viscosities calculated
 for  hadron mixture at  $n_{\rm bar}=n_0$ and  $4n_0$ (solid lines).
 Calculations performed in the IG based model
with the same hadron set as in the SHMC model are demonstrated by
dashed lines.
 }
 \label{eta_nB}
\end{figure}

 For the case of the multi-component hadron
mixture within IG and SHMC models the temperature dependence of
the reduced $T^4$-scaled shear and bulk viscosities are shown at
baryon densities $n_{\rm bar}=n_0$ and  $4n_0$ ($n_0$ is the
nuclear saturation density)  in the left and right panels of
Fig.~\ref{eta_nB},
respectively. The reduced shear viscosity  calculated in the SHMC
model (solid lines)  is close to that in the IG model with the
same hadron set (dashed lines). Differences in the $\eta/(\tau
T^4)$ ratio for the IG and SHMC models   appear only at high
temperatures $T\gsim 150$ MeV. At $T\lsim  100$ MeV the reduced
$T^4$-scaled bulk viscosity (right panel) in the IG based model
proved to be  larger than  that in the SHMC model. Contrary, for
larger $T$ the reduced bulk viscosity  in the IG model becomes
smaller than that in the SHMC model. Differences come
 from the strong dependence of the bulk viscosity $\zeta$
on the values of thermodynamical quantities (see
Eqs.(\ref{Qa}),(\ref{bulk})).  Note that at
$T\gsim$100 MeV and  $n_{\rm bar}\gsim n_0$ the shear and
bulk viscosities are getting comparable in magnitude. Growth of
the relative importance of $\zeta$ with increase of temperature
seems to be quite natural because the bulk viscosity takes into
account
 momentum dissipation due to inelastic channels which number
increases with the temperature increase.

\subsection{Estimation of the relaxation time}

The relaxation time is defined by Eq.~(\ref{tau}). We implement
free cross sections  in case of the IG based model, similar to
procedure performed in Ref.~\cite{PPVW93}. In case of the SHMC
model,  the in-medium modification of cross sections is
incorporated by a shift of a ``pole'' of the collision energy by
the mass difference $m_a-m_a^*$ according to prescription of
Ref.~\cite{BC08}. Due to a lack of microscopic calculations this
is the only modification which we do here.
 Important peculiarity of the nucleon
contribution to the relaxation time at low temperature is
associated with the particular role played by the Pauli blocking.
It means that appropriate multi-dimensional integration should be
carried out quite accurately with using quantum statistical
distribution functions. Calculations using the kinetic
Uehling-Uhlenbeck equations  for the purely nucleon system in the
non-relativistic approximation were performed in~\cite{Dan84}. For
$T\lsim 100$ MeV an extrapolation expression has been obtained:
\be
{\tilde\tau}_{NN} \  &\simeq& \frac{850}{T^2} \left(\frac{n_{\rm
bar}}{n_0}\right)^{1/3} \ \left[1+0.04T \frac{n_{\rm
bar}}{n_0}\right] + \frac{38}{T^{1/2}(1+160/T^2)}\frac{n_0}{n_{\rm
bar}}~. \label{apptauN}
 \ee
 Thus the
relaxation time demonstrates well known behavior   $T^{-2}$, for
$T\to 0$.

 To simplify calculations
we  use  Eq. (\ref{apptauN}) for the partial nucleon relaxation
time $\tilde\tau_{NN}$, to be valid at low temperatures,  smoothly
matching it (at $T\sim 100$ MeV)  with the partial nucleon
contribution calculated following Eq. (\ref{tau})  for higher
temperatures. We take into account the whole hadron set involved
into the SHMC model. The relaxation time for every
component is evaluated according to Eq. (\ref{tau}).

\subsection{Collisional viscosity  in  heavy ion collisions}

Above we have studied reduced viscosities of the hadron matter at
different temperatures and baryon densities. In reality a hot and
dense system being formed in a heavy-ion collision then expands
towards freeze-out state, at which the components stop to interact
with each other. Here we use the freeze-out curve $T_{\rm
fr}(\mu_{\rm bar}^{\rm fr})$ extracted from analysis of
experimental particle ratios in statistical model for many species
at the given collision energy $s_{NN}^{1/2}$ treating  the freeze-out
temperature $T_{\rm fr}$ and chemical potential $\mu_{\rm
bar}^{\rm fr}$ as free parameters~\cite{COR06,ABMS06}.

\begin{figure}[thb]
 \hspace*{15mm}
\includegraphics[width=90mm,clip]{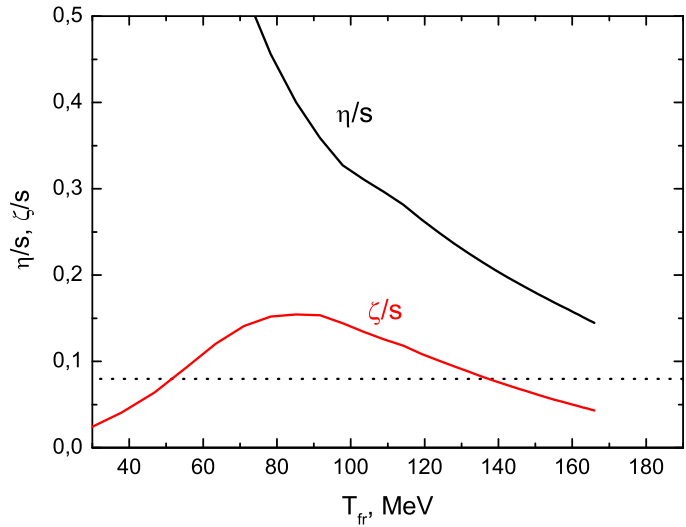}
\caption{ Shear and bulk viscosities per entropy density
calculated in the SHMC model  for central Au+Au collisions along
the freeze-out curve (at $T=T_{\rm fr}$) ~\cite{COR06} for the
baryon enriched system. The dotted line is the lower AdS/CFT bound
$\eta/s=1/4\pi$~\cite{KSS03}.
 }
 \label{sh_b_fr}
\end{figure}

 In Fig.~\ref{sh_b_fr}, viscosity coefficients per entropy
density $s$ are shown versus the freeze-out temperature for Au $+$
Au collisions (which  is unambiguously related to the freeze-out
chemical  potential $\mu_{\rm bar}^{\rm fr}$~\cite{COR06} needed
to calculate thermodynamical quantities at the freeze-out).
Dimensionless ratios of the viscosity to the entropy density
$\eta/s$ and $\zeta/s$ characterize the energy dissipation  in the
medium. As  we see,  the $\eta/s$ ratio decreases monotonously
with increase of the temperature, being higher than the lower
bound $1/4\pi$ but tending to it with further increase
of $T_{\rm fr}$.  The value $\zeta/s$ exhibits a
maximum at $T_{\rm fr}\sim 85$ MeV and then tends to zero with
subsequent increase of $T_{\rm fr}$. As has been emphasized above,
at $T\gsim$100 MeV values of the shear and bulk viscosities become
quite comparable, $(\eta/s)_{\rm fr}\simeq 2(\zeta/s)_{\rm fr}$.
\begin{figure}[thb]
 \hspace*{10mm}
\includegraphics[width=90mm,clip]{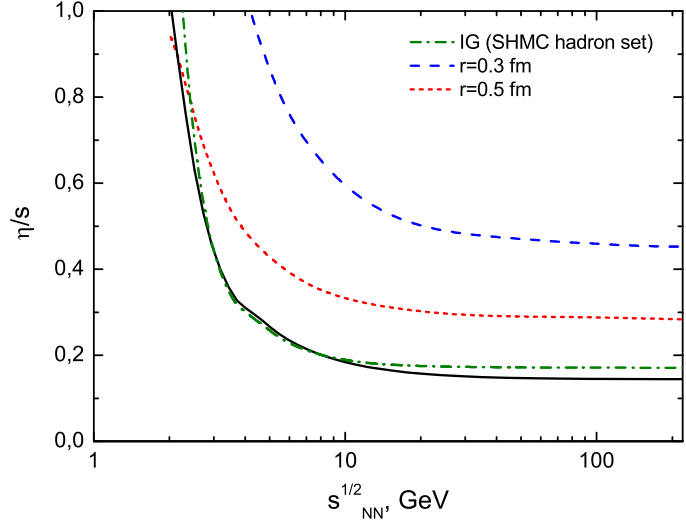}
\caption{The  ratio of the shear  viscosity to the entropy
density calculated for central Au+Au collisions along the chemical
freeze-out curve~\cite{COR06} within the SHMC model as a function
of the collision energy $s^{1/2}_{NN}$ (solid line). Dashed and
short-dashed curves are the results of the excluded volume hadron
gas model~\cite{GHM08} with hard-core radii $r=$0.3 and $r=$0.5
fm, respectively. The dot-dashed line corresponds to the  IG
model with the same set of hadrons as for the SHMC
model.  }
 \label{sh_Gfr}
\end{figure}

 In Fig. \ref{sh_Gfr}, the $\eta/s$ ratio calculated in our SHMC
model  (solid line) is plotted as a function of the collision
energy $\sqrt{s_{NN}}$ of two  Au+Au nuclei. The result for the IG
model with the same hadron set as in SHMC model  is plotted by the
dash-dotted line. We  note that for $\sqrt{s_{NN}}\gsim 3$ the
SHMC results prove to be very close to the IG based model ones
(with the same hadron set as in SHMC model), since the freeze-out
density is rather small and the decrease of the hadron masses
occurring in the SHMC model is not important. The results for  the
hadron hard core gas model (the van der Waals excluded volume
model)~\cite{GHM08} at two values of the particle hard core radius
$r$  are shown by dashed and short-dashed lines. In all cases for
$\sqrt{s_{NN}}\gsim 2$ GeV the ratio $\eta/s$ decreases along the
chemical freeze-out line with increasing the collision energy and then
flattens at $\sqrt{s_{NN}}\gsim $10 GeV, since  freeze-out at such
high collision energies already occurs at almost constant value of
$T_{\rm fr}\approx 165$ MeV. The shear viscosity  of the
non-relativistic Boltzmann gas of hard-core particles~\cite{GHM08} is
$\propto{\sqrt{mT}}/{r^2}$.
Since Fermi statistical effects are not included within this
model, the shear viscosity, $\eta$, decreases with decrease of
$T$. Nevertheless the $\eta/s$ ratio increases and diverges at low
energy/temperature, as the consequence of a more sharp decrease of
the entropy density compared to $\eta$, see Fig.~\ref{sh_Gfr}. As
follows from the figure,  the smaller $r$ is, the higher $\eta/s$  is
in the given excluded volume model.
For $\sqrt{s_{NN}}\gsim 4$ and $r\simeq 0.7$ fm the $\eta/s$ ratio
 is expected  to be close to the values computed in the IG and SHMC models.

Recently an interesting attempt has been undertaken in
~\cite{IMRS09} to extract the shear viscosity from the 3-fluid
hydrodynamical analysis of the elliptic flow in the AGS-SPS energy
range. An overestimation of experimental $v_2$ values in this
model was associated with dissipative effects occurring during the
expansion and freeze-out stages of participant matter evolution.
 The resulting
values of $\eta/s$  vary in  interval $\eta/s\sim 1-2$ in the
considered domain of $\sqrt{s_{NN}}\approx 4-17$ GeV
(corresponding to temperatures $T\approx 100-115$
MeV)~\cite{IMRS09}.  Authors consider their result as an upper
bound  on the $\eta/s$ ratio in the given energy range.
 Note that mentioned values are much higher than
those which follow from our estimations given above and presented
in Figs. \ref{sh_b_fr} and \ref{sh_Gfr}.

Other microscopic estimate of the share viscosity to the entropy
density ratio for the relativistic hadron  gas based on the UrQMD
code was performed in Ref.~\cite{DB08} where 55 baryon species and
their antiparticles and 32 meson species were included. The full
kinetic and chemical equilibrium is achieved at $T=$130 and 160
MeV, respectively. The extracted ratio $\eta/s\gsim$1 exceeds the
SHMC result by a factor of  5. Introducing a non-unit fugacity or
a finite baryon density allows one to decrease the ratio twice but
nevertheless it is still too high as compared to both the SHMC
result and  the lower bound $\eta/s=1/4\pi$. Analyzing their
result  authors~\cite{DB08} conclude that the dynamics of the
evolution of a collision at RHIC is  dominated by the deconfined
phase (exhibiting very low values of $\eta/s$) rather than by the
hadron phase. Note however that in-medium effects in the hadron
phase are not included into consideration in the UrQMD model
though, namely, these effects result in the required decrease of
the $\eta/s$ ratio in our SHMC model.

\section{Conclusions}\label{Concl}

 In this paper, we derived expressions for the shear and bulk
viscosities in the relaxation-time approximation for a hadron
system described by the quasiparticle relativistic mean-field
theory with  scaling of hadron masses and couplings (SHMC). The
EoS of the SHMC model fairly well reproduces global properties of
hot and dense  hadron matter including the temperature region near
$T_c$ provided all coupling constants $g_{\sigma b}$ are strongly
suppressed except for  nucleons. Thus obtained kinetic
coefficients are compared with those calculated in other models of
the hadron matter.

 With increasing freeze-out temperature $T_{\rm fr}$  (for central
 Au+Au collisions), the
$\eta/s$ ratio undergoes a monotonous decrease approaching values
close to the AdS/CFT  bound at $T\sim T_c $ MeV, while the
$\zeta/s$ ratio exhibits a maximum at $T_{\rm fr}\sim$85 MeV. In a
broad temperature interval the $\eta/s$ and $\zeta/s$ ratios are
not small and viscous effects can be noticeable. The viscosity
values at the freeze-out can be transformed into  dependence on the
colliding energy $\sqrt{s_{NN}}$ (for central Au+Au collisions).
When the collision energy decreases, the $\eta/s$ goes up.  The
high-energy flattening of the $\sqrt{s_{NN}}$ dependence occurs at
quite low $\eta/s<0.2$. It implies that a small value of $\eta/s$
required for explaining a large elliptic flow observed at RHIC
could be reached in the hadronic phase. This might be an important
observation which we have demonstrated within the SHMC model.

The $v_2$ analysis  indicates to  different values of
$\eta/s$ for peripheral and  central collisions. Therefore,  it
would be interesting to perform hydrodynamic calculations
using the $T-\mu_{\rm bar}$ dependent transport coefficients rather
than constant ones. The need of such an approach was recently
emphasized in \cite{Cha09}. Further we will use the SHMC model EoS
 with the derived transport coefficients for this purpose.

\vspace*{5mm} {\bf Acknowledgements} \vspace*{5mm}

We are grateful to  K.K.~Gudima, Y.B.~Ivanov, Y.L.~Kalinovsky and
 E.E.~Kolomeitsev for numerous  discussions and valuable remarks.  This work was
supported in part by the  BMBF/WTZ project  RUS 08/038, the RFFI grants 
08-02-01003-a and 10-02-91333 ммхн-Ю, the Ukrainian-RFFI grant 
╧ 09-02-90423-СЙП-Т-a, the DFG grant WA 431/8-1 and the Heisenberg-Landau grant.

\newpage
\vspace*{3cm}
\begin{center}
\Large юММНРЮЖХЪ \\[4mm]
{\bf бЪГЙНЯРЭ ЮДПНММНИ ЛЮРЕПХХ Б ПЕКЪРХБХЯРЯЙНИ ЛНДЕКХ ЯПЕДМЕЦН ОНКЪ
ЯН ЯЙЕИКХМЦНЛ ЮДПНММШУ ЛЮЯЯ Х ЙНМЯРЮМР ЯБЪГХ}\\[5mm]
ю.я. уБНПНЯРСУХМ, б.д. рНМЕЕБ Х д.м. бНЯЙПЕЯЕМЯЙХИ\\[2mm]
\end{center}
\large
яДБХЦНБЮЪ ($\eta$) Х НАЗЕЛМЮЪ($\zeta$)  БЪГЙНЯРХ БШВХЯКЪЧРЯЪ Б ЙБЮГХВЮЯРХВМНЛ 
ОПХАКХФЕМХХ БПЕЛЕМХ ПЕКЮЙЯЮЖХХ ДКЪ ЮДПНММНИ ЛЮРЕПХХ, НОХЯШБЮЕЛНИ Б ПЮЛЙЮУ
ПЕКЪРХБХЯРЯЙНИ ЯПЕДМЕОНКЕБНИ ЛНДЕКХ ЯН ЯЙЕИКХМЦНЛ ЮДПНММШУ ЛЮЯЯ Х ЙНМЯРЮМР 
ЯБЪГХ. оПЕДЯРЮБКЕМН ЯПЮБМЕМХЕ Я ПЕГСКЭРЮРЮЛХ ДПСЦХУ ЛНДЕКЕИ. оНЙЮГЮМН, ВРН 
ЛЮКНЕ ГМЮВЕМХЕ НРМНЬЕМХЪ ЯДБХЦНБНИ БЪГЙНЯРХ Й ОКНРМНЯРХ ЩМРПНОХХ, РПЕАСЕЛНЕ
ДКЪ НАЗЪЯМЕМХЪ АНКЭЬНЦН ЩККХОРХВЕЯЙНЦН ОНРНЙЮ, МЮАКЧДЮЕЛНЦН Б ЩЙЯОЕПХЛЕМРЮУ
МЮ RHIC, ЛНФЕР АШРЭ ДНЯРХЦМСРН Б ЮДПНММНИ ТЮГЕ. нРЛЕВЮЧРЯЪ ЯПЮБМХРЕКЭМН 
АНКЭЬХЕ ГМЮВЕМХЪ НАЗЕЛМНИ БЪГЙНЯРХ  Б ЯКСВЮЕ АЮПХНМН-НАНЦЮЫЕММНИ ЛЮРЕПХХ.  

\end{document}